# Revolutionizing Wireless Communication with Single Layer Capacitor-Based Antenna Technology

Alex Nguyo Gachahi

Senior Technician, School of Physics and Earth Sciences, Technical university of Kenya

Corresponding address: Alex N. Gachahi  email: agachahi@tukenya.ac.ke

**List of Abbreviations:**

| | |
|---|---|
| **DF** | Dissipative Factor |
| **ESR** | Equivalent Series Resistance |
| **FCC** | Federal Communications Commission |
| **GPS** | Global Positioning System |
| **GSM** | Global System for Mobile Communications |
| **IoT** | Internet of Things |
| **ITU** | International Telecommunication Union |
| **LTE** | Long-Term Evolution |
| **NFC** | Near Field Communication |
| **PCB** | Printed Circuit Board |
| **RF** | Radio Frequency |
| **RSSI** | Received Signal Strength Indicator |
| **V2X** | Vehicle-to-Everything |
| **VSWR** | Voltage Standing Wave Ratio |

# 1. Introduction

## Unleashing the Potential of Wireless Communication

In an era where connectivity defines how we live, work, and interact, the technology underpinning our wireless networks has never been more critical. As we push the boundaries of what's possible with the Internet of Things (IoT), telecommunication, healthcare innovations, and autonomous vehicles, the demand for more efficient, reliable, and compact wireless communication solutions escalates. The advent of Single Layer Capacitor-Based Antenna Technology marks a significant leap forward. This patented technology addresses the core challenges that have long constrained the evolution of wireless systems. This white paper delves into a groundbreaking approach that not only promises to revolutionize antenna design but also paves the way for a new wave of connectivity solutions. By offering unprecedented efficiency, compactness, and versatility, this technology stands to redefine the landscape of wireless communication, making it more accessible, robust, and adaptable to the needs of our rapidly evolving digital society.

# 2. Executive Summary

## Transforming Connectivity with Single Layer Capacitor-Based Antenna Technology

In the quest for seamless wireless connectivity, the Single Layer Capacitor-Based Antenna Technology emerges as a game-changer, offering a suite of advantages that propel it far beyond the capabilities of traditional antenna designs. Engineered to meet the growing demands for higher efficiency, smaller form factors, and greater flexibility, this technology heralds a new era in wireless communication systems.

**Key Innovations and Benefits:**

i. **Compactness:** With a minimal footprint, these antennas enable the development of smaller, more discreet devices without compromising on performance.

ii. **High Efficiency:** Achieving an efficiency rate of 97%-99%, the technology ensures superior signal strength and extended battery life for battery-powered devices.

iii. **Broad Frequency Range:** Capable of operating across a wide spectrum from few khzs to 100 GHz, this is due to the fact that base elements adaptable to this frequency range are already available in the market today. This technology offers unparalleled versatility, supporting everything from IoT applications to advanced telecommunication networks.

iv. **Wide bandwidth:** the behavior of single layer capacitors around their resonant frequencies ensures a very wide bandwidth. This is explained further in the technical data section.

v. **Cost-Effectiveness:** the base element in this technology is the single layer capacitor, these elements, compatible with most existing wireless technologies, are already in the market at very competitive prices as compared to traditional antenna.

Single Layer Capacitor-Based Antenna Technology not only meets the technical demands of modern wireless communication but also addresses the economic and spatial challenges faced by device manufacturers and network providers. By delivering high performance in a smaller package at a lower cost, it enables a broad range of applications — from enhancing mobile device connectivity to enabling robust IoT ecosystems and smart city infrastructures. This white paper outlines the technology's technical underpinnings, showcases its wide applicability, and guides on its integration into existing systems, illustrating its potential to reshape wireless communication landscapes globally.

## 3. Technology Overview

### Pioneering Advances with Single Layer Capacitor-Based Antenna Technology

At the heart of the Single Layer Capacitor-Based Antenna Technology lies a fundamental reimagining of antenna design, leveraging the simplicity and efficiency of single-layer capacitors. This section explores the innovative mechanics behind the technology, its distinguishing features, and the technical specifications that set it apart from traditional antenna solutions.

The Mechanics:

Traditional antennas, constrained by size and efficiency limitations, often strive to meet the demands of modern wireless communication. Our technology, however, utilizes single-layer capacitors as the antenna base elements in an array or single element configuration to overcome these challenges. Unlike conventional antennas that rely on physical dimensions to determine frequency response (spatial resonance), our design exploits the electrical properties of single layer capacitors. Unlike traditional antennas, single layer capacitor based antennas rely on temporal resonance eliminating the dependence of traditional antennas on spatial resonance. This allows for a compact size without sacrificing performance across a broad frequency range.

Unique Features:

i. **Compact Design:** The use of single-layer capacitors significantly reduces the antenna's physical footprint, single layer capacitors with specifications advanced enough to be adaptable to all major wireless systems, an example of which is presented later in the technical data section, are already available in the market. These capacitors have a cuboid structure of less than a millimeter in length, width and thickness; for example, the single layer capacitor element whose data analysis is presented in the technical data section has a cuboid structure measuring 0.63mm by 0.63mm by 0.25mm. This level of compactness not only reduces the footprint of the antenna but also opens up the versatility of antennas designed for specific applications by taking advantage of antenna arrays and antenna diversity methods that were not accessible in traditional antennas. This makes it ideal for integration into a wide array of devices, from smartphones to IoT sensors with unprecedented versatility.

ii. **Enhanced Efficiency**: as shown later in the technical data section single layer capacitors are highly efficient with efficiencies of between 97%-99%. The efficiency of a single layer capacitor can be extracted directly from the data sheet in form of the dissipative factor. The dissipative factor (DF) of a capacitor is the percentage ratio of the energy lost due to its equivalent series resistance (ESR) and dielectric losses, to the reactive power supplied at the operating frequency. The DF of a capacitor can also be extracted from the s-parameter data of a capacitor as shown in the technical data section. The value of DF for most capacitors is less than 3%. This level of efficiency ensures robust signal transmission and reception, minimizing energy loss and optimizing device performance.

iii. **Versatile Frequency Range:** single layer capacitors adaptable to wireless systems operating at frequencies from a few kilohertz to up to 100 GHz are commercially available in the market today, which is virtually all communication systems in use today; examples of these systems are 5G networks, Wi-Fi, Bluetooth, and satellite communications.

iv. **Wide bandwidth:** The impedance of single layer capacitors remains low and stable for a large frequency range around the resonant frequency as shown in the technical data analysis. This results in an ultra-wide bandwidth. The ESR of the capacitor also remains fairly stable over this frequency range as well. This results in the capacitor remaining highly efficient across this frequency range.

v. **Cost-Effectiveness:** Single layer capacitors which are the base elements of our antenna cost between 0.1 dollars for low frequency capacitors to about 10 dollars for the high frequency capacitors operating in the 10s of gigahertz frequencies. For the most common and mostly used communication systems like Wi-Fi, Bluetooth, GSM and satellite communication, the single layer capacitors in the market fitting their frequency range which is between 0.1ghz to 20ghz, cost about 0.7 dollars per element, the base element whose data is presented in the technical data costs 0.75 dollars and is adaptable to systems operating in the frequency range 0.5ghz to 6ghz. As this technology is adopted across the wireless communication industries, the cost of the base elements is also expected to drop owing to economies of scale. This will enable more affordable and versatile wireless solutions.

**Advantages of the novel antenna over the existing antennas**

The single-layer capacitor-based antenna family (Novel antenna) offers unique advantages across industries and applications. Its wide frequency range, compact size, high efficiency, and distinctive radiation patterns make it a promising solution for wireless communication systems, broadcasting, wireless data transmission, aerospace and defense, automotive applications, public safety, medical devices, research, and scientific applications. By leveraging these advantages and considering specific market requirements, the single-layer capacitor-based antenna family presents a significant opportunity for innovation and advancements in wireless communication technology. Depending on the target market and existing technologies, these benefits contribute to improved signal reception, extended coverage, enhanced communication capabilities, and reduced power consumption.

**Challenges of the novel antenna as compared to existing antennas**

During our research the most prominent challenge identified in relation to the novel antenna is its impedance profile. The single layer capacitor-based antenna has very low impedance around the capacitor's resonant frequency. this results in an impedance mismatch when integrated into existing wireless networks considering most wireless networks are designed with around 50Ω antenna input/output pins. However, this can be mitigated by including impedance matching mechanisms. An example impedance matching mechanism is introduced later in the simulation results and technical data analysis section.

## 4. Simulation results and technical data analysis

The directivity, gain and the radiation patterns of an antenna build using single layer capacitors are obtained by solving Maxwell's equations in space around the spacemen as shown in Figures 1-3 below. These simulations were done using MATLAB for a single element and for multiple elements together to simulate an antenna array

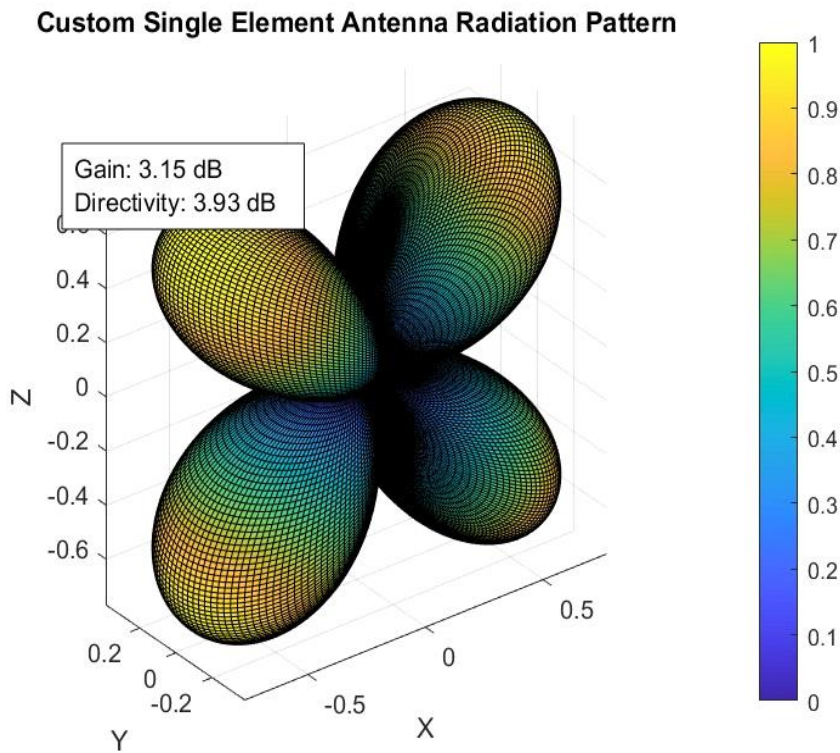

**Figure 1:** The radiation pattern of a single element of the novel antenna showing the four main lobes of this antenna

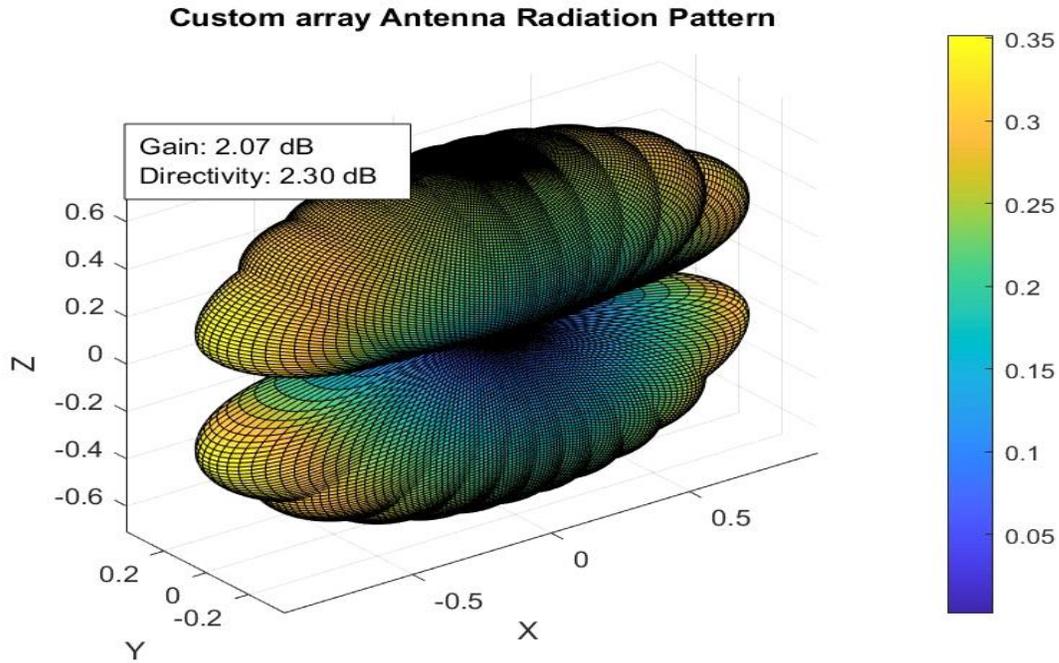

**Figure 2:** The nearfield radiation pattern of an array made up of the Novel Antenna elements. It shows the four main lobes and also the minor lobes

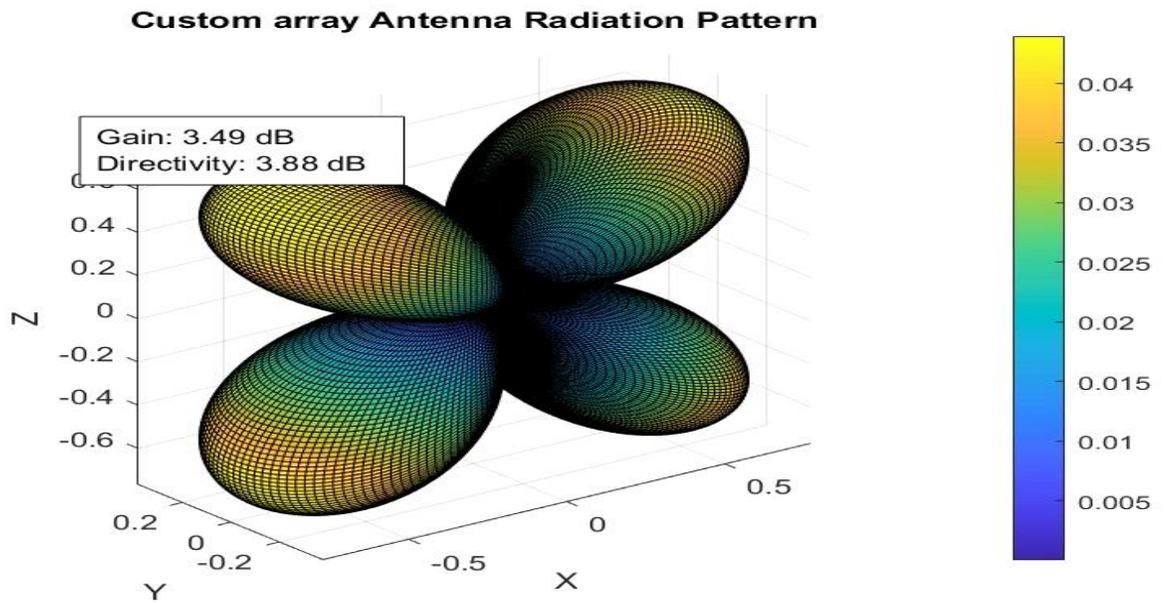

**Figure 3: The far field radiation pattern of an array made up of Novel Antenna elements showing the four main lobes**

### Technical data analysis

In order to study the compatibility with existing systems, and the unique properties attributed to this antenna, we analyzed s-parameter data of a single layer capacitor in the market today using MATLAB. The analysis whose results are presented here show that our antenna though an impedance matching mechanism is necessary, our antenna is easily integrated into existing systems as constituted (in terms of their impedance profiles) at minimum cost in terms of design changes or materials. For an advanced proof of concept, we integrated the single layer capacitor presented here into a sim800l *gsm* chip from *Simcom*. The data obtained is also presented here. From this analysis a general impedance matching mechanism is proposed.

S-parameter data analysis results

The overall impedance of the antenna is very low, ranging from less than 0.5 to 3 ohms across the frequency range of 0.1 GHz to 20 GHz. This is significantly lower than the typical impedance of 50 ohms seen in standard antennas and transmission lines. The low impedance of our antenna as compared to typical impedance of existing systems results in an impedance mismatch between our antenna and the system in question. The impedance curve is relatively smooth and flat, indicating a consistent response across the frequency range (Figure 5). This is in contrast to traditional antennas, which often exhibit peaks and valleys in their impedance. The resistance values are also relatively low, ranging from about 0.5 to 1.5 ohms across the frequency range. This further contributes to the low overall impedance of the antenna, and shows the efficiency of the antenna across the frequency range.

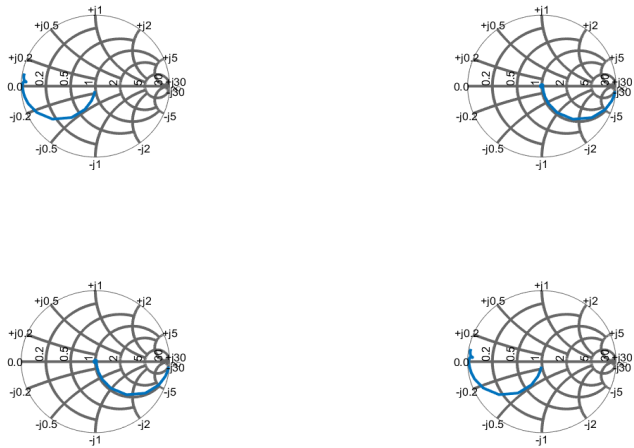

**Figure 4: The s-parameter data visualization of-from top left moving clockwise- s11,s12,s21 and s22 parameters; note that the real part of the elements impedance is low and remains relatively unchanged over the frequency range under study. The reactive part of the impedance and hence the impedance is also very low and increases steadily as you move away from the element's resonant frequency.**

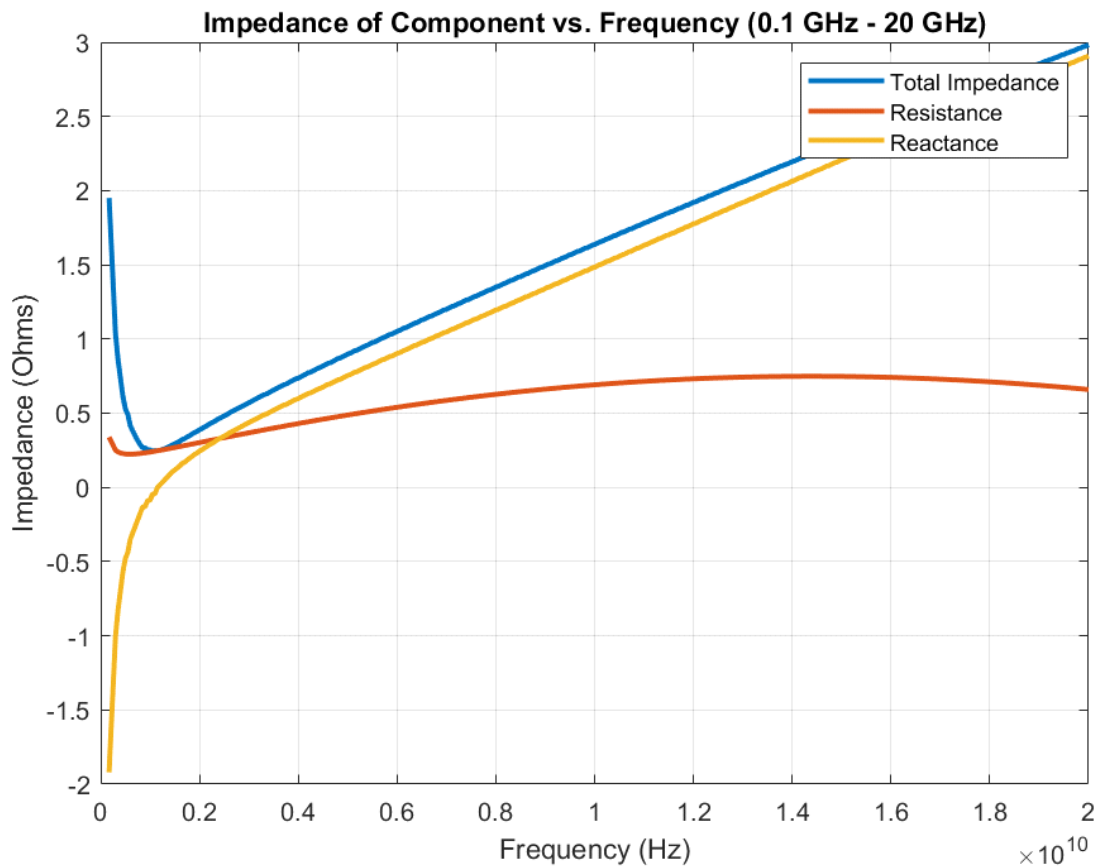

**Figure 5: A graph of the frequency response of impedance, reactance and resistance of the antenna note: the resistance remains low over the frequency range with only a slight variation. The reactance and the impedance show a steady increase over the frequency range but overall remains low (below 3Ω). The resonant point of the element is at the frequency which the reactance is 0 and impedance is minimum.**

The plot of the dissipative factor vs. frequency shows how losses within the capacitor change with frequency. The DF shows sharp increases near the upper and lower frequency edges but remains below 2% for most of the frequency range (Figure 6). This information is critical for understanding the efficiency of the antenna across its operating bandwidth.

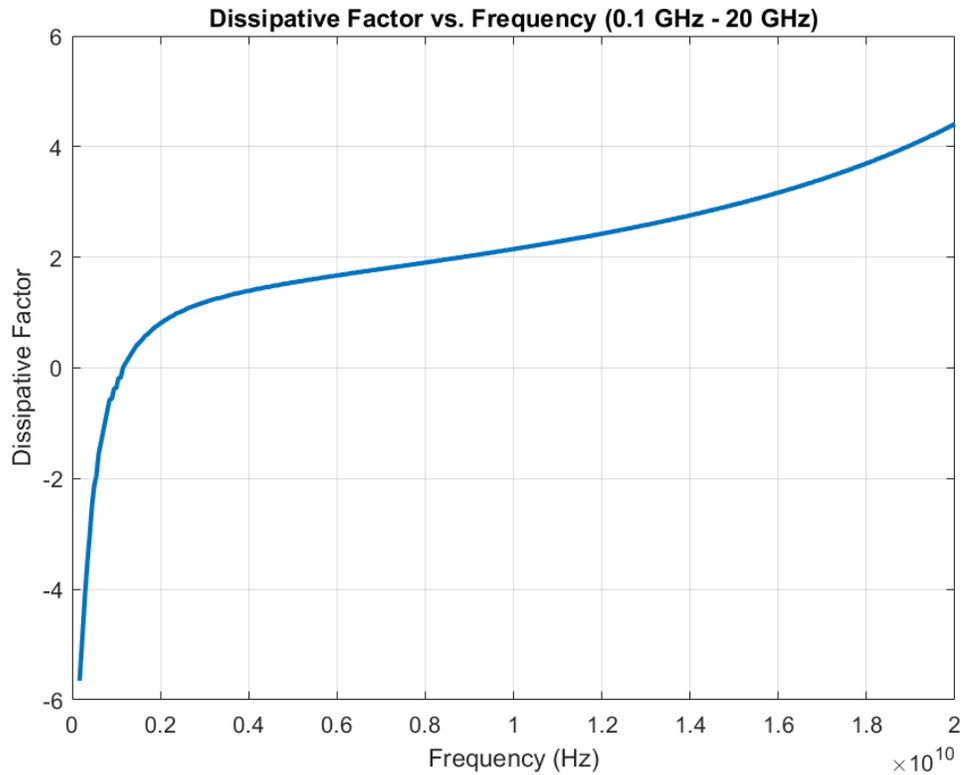

**Figure 6: A graph of the frequency response of the dissipative factor of the base element. Note: the dissipative factor, which is a measure of efficiency in single layer capacitors, increases sharply around the edges but overall remains below 6%.**

*Sim800l* single layer capacitor-based antenna integration data

In an advanced proof-of-concept experiment, a novel antenna based on single-layer capacitors was integrated into a *SIM800L GSM* module by *SimCom*. This integration was achieved by substituting the module's antenna connector with the single-layer capacitor whose s-parameter data is analyzed above. A second SIM800L module was equipped with a conventional micro-strip antenna for comparative analysis. The evaluation involved collecting data on the signal strengths of each module in three distinct environments, each chosen for their different levels of expected signal coverage; Utilizing the *SIM800L's* internal *AT+CSQ* command to report Received Signal Strength Indicator (RSSI) with values ranging from 0 to 31.The first environment was an urban center, anticipated to have robust coverage, where the average RSSI value for the module with the novel antenna was 20, corresponding to a signal strength of approximately -73 dBm; indicating a good signal quality. In contrast, the micro-strip antenna-equipped module recorded an RSSI of 23, or about -67 dBm, suggesting slightly better signal reception. In the semi-urban area, expected to have moderate coverage, the novel antenna's RSSI was 11(approximately -89 dBm), while the micro-strip antenna recorded an RSSI of 15(approximately -83 dBm). In the rural location, presumed to have weak signal coverage, the novel antenna achieved an RSSI of 10, translating to a signal strength of around -93 dBm, in contrast to 13 (approximately -89 dBm) for the micro-strip antenna. Across all locations, the performance of the novel antenna reached approximately 70%- 80% of that observed with the micro-strip antenna, when considering the RSSI values in the context of their dBm equivalents. This diminished signal strength exhibited by our antenna compared to the microstrip antenna is attributed to the impedance mismatch between the SIM800L module's antenna input and the innovative antenna. Following this analysis, an impedance matching network was proposed to address this discrepancy. The results from the simulations of this proposed impedance matching mechanism are outlined in the next section.

figure7 is a representation of the signal strength against time for the two setups in the semi-urban area. dataset1 represents data from the setup incorporating a micro strip antenna while dataset2(in blue) represent the data from our novel antenna. The percentage difference between the two means is calculated to be 26.34%. A statistical significance analysis of the two means was carried out on these two sets. The p-value obtained, p = 0.000, shows the difference between the means is statistically significant. This result shows that we were able to obtain a 74% performance of our antenna as compared to the microstrip antenna utilizing an area that is about 600 times smaller than that of the microstrip (the novel antenna occupies an area of $0.3969mm^2$ and the microstrip occupies an area of $245mm^2$). It's also important to note that in this case no impedance matching network is included. With an impedance matching network included the performance of the novel antenna is expected to surpass that of the microstrip antenna.

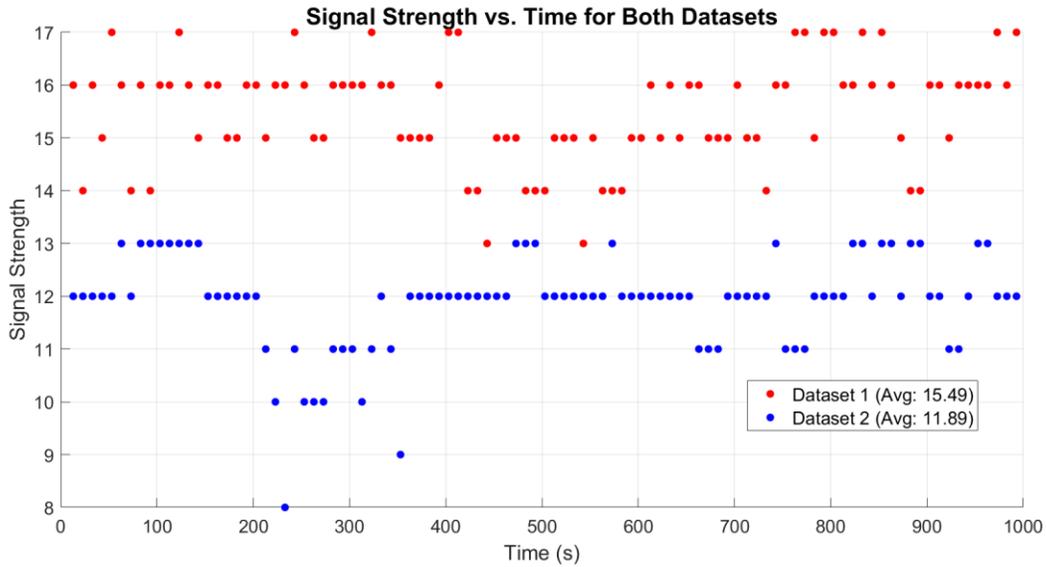

**Figure 7:** A graph of signal strength against time for the novel antenna (dataset2) and the micro strip antenna (dataset 1) note: the micro-strip antenna shows an overall higher signal strength than the novel antenna. This is attributed to the impedance mismatch between the module and the novel antenna.

### Impedance matching

The challenge of impedance mismatch presented a significant hurdle in the initial deployment of our antenna technology. A mismatch between the antenna's low impedance and the standard 50-ohm input/output pins of most communication systems was observed, leading to suboptimal power transfer and reduced signal quality. To address this, we embarked on a design and simulation process, resulting in the development of a purely resistive impedance matching network. This network, tailored to seamlessly integrate with the 50-ohm standard, is the most general network from which other specialized networks can evolve. This method not only resolved the impedance mismatch but also maintained the antenna's wide bandwidth. Figure 8 and Figure 9 are graphs of impedance vs frequency and voltage standing wave ratio (VSWR) against frequency respectively. The Voltage Standing Wave Ratio (VSWR) for the integrated novel antenna was initially simulated to be about 50, indicating a significant impedance mismatch and inefficient power transfer from the SIM800L module to the antenna, which could have contributed to the lower signal strengths observed. However, with the incorporation of an impedance matching network, the VSWR improved dramatically to 1.04. This near-ideal VSWR value signifies almost perfect impedance matching, ensuring that the vast majority of the power is effectively radiated by the antenna with minimal reflections. This optimization dramatically enhances the antenna's performance, potentially improving signal strength and reliability across the tested environments and narrowing the performance gap noted between the novel antenna and the conventional micro-strip antenna. This improvement in impedance matching and the subsequent enhancement in VSWR are critical for the antenna's efficiency and the overall performance of the *SIM800L* module in real-world applications. By effectively addressing the primary technical challenge of impedance mismatch, the novel antenna design, augmented with the impedance matching network, presents a compelling advancement for GSM and broader wireless communication applications, optimizing power usage and enhancing device performance.

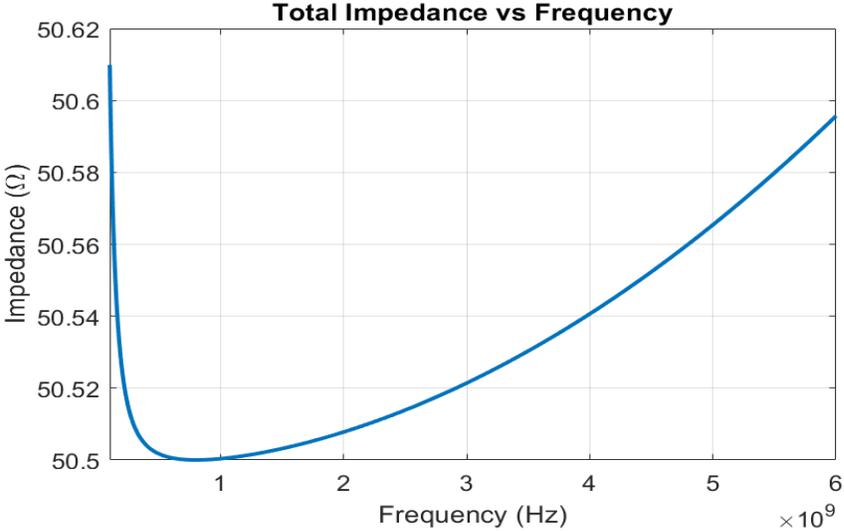

Figure 8: A graph of impedance vs frequency for a 50ohm source matched to the novel antenna using a resistor: note that the impedance remains within a very narrow range around 50Ω over the frequency range.

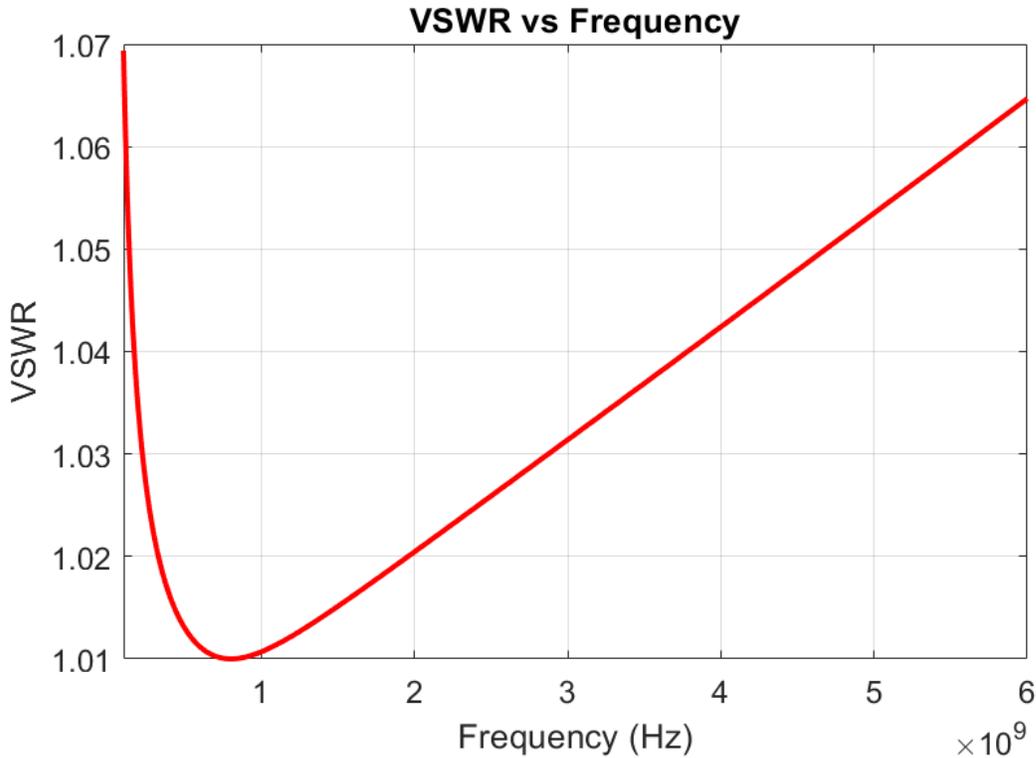

**Figure 9: A graph of VSWR vs frequency for a 50ohm source matched to the novel antenna using a resistor. Note the voltage standing wave ratio (VSWR) remains below the maximum recommended VSWR value of 2 over the frequency range.**

This technological breakthrough not only addresses the immediate challenges faced by the wireless communication industry but also opens up new avenues for innovation and application development. By providing a scalable, efficient, and versatile antenna solution, Single Layer Capacitor-Based Antenna Technology is poised to play a crucial role in the future of connectivity.

## 5. Market Potential and Analysis

## Catalyzing Innovation across Industries

The Single Layer Capacitor-Based Antenna Technology stands at the precipice of transforming connectivity in numerous sectors. Its versatility, efficiency, and compact size align with the needs of a broad spectrum of applications in today's market. This section delves into the technology's potential impact across key industries, supported by specific current market applications.

## Telecommunications:

i. **Mobile Networks**: These include 5g, 4g and 3g networks. With its broad frequency range and high efficiency, the technology is ideal for mobile network base stations and devices, facilitating faster data speeds and more reliable connections; for example, the biggest challenge in actualizing 5G networks is the effective range of the technology, this can be attributed to higher levels of attenuation by the atmosphere as compared to lower frequency mobile networks such as the 4G. However, the technology presented here overcomes these challenges by allowing the design of smaller elements that can be put in to arrays. These configurations can take advantage of antenna diversity techniques to enhance the performance of the antennas. The low impedance of the base elements also enhances the sensitivity of the antennas. The higher efficiencies also improve the overall efficiency of client devices resulting in longer battery lives.

ii. **Satellite Communications**: includes communication with satellites, enabling services like global positioning systems (GPS) and satellite internet, making it vital for remote connectivity solutions. The antennas required for communication with satellites are bulky and limited to highly immobile setups, this has hampered the adoption of this technology into more compact devices. This makes mobile communication dependent on terrestrial systems which are expensive. By utilizing the advantages offered by our novel antenna in terms of size, efficiency and overall versatility, satellite communication may finally break the barrier allowing integration into our personal devices like mobile phones. This would greatly enhance communication even in remote areas where terrestrial systems don't reach.

## Consumer Electronics:

i. **Smartphones and Wearables**: By reducing antenna size without compromising performance, the technology enables sleeker designs and longer battery life in smartphones, smartwatches, and fitness trackers. With the advent of cutting edge technologies in this industry for example virtual reality, the need for more robust communication solutions is necessary to keep up with the requirements of these technologies. By designing these systems from scratch to use the Single Layer Capacitor Based Antenna Technology, more robust solutions will now be available leading to unprecedented growth in the capabilities of these systems.

ii. **IoT Devices**: Powers a multitude of IoT applications, from smart home devices (e.g., security systems, thermostats) to industrial IoT sensors. These technologies make use of wireless data transmission technologies like Wi-Fi, Bluetooth, NFC technologies, Lora and others. By taking advantage of our novel antennas capabilities, several constrains in this sector can be overcome e.g. the range of the IoT devices can be greatly increased. Adoption of this technology will result in higher rates of data transfer between the devices leading to more versatile solutions. This will open up the solution space around the IoT sector.

### Broadcasting (television, radio)

This sector utilizes lower frequencies than other communication systems. This results in bulky antennas for client devices that are often mounted outside. Taking advantage of the novel antennas unique selling points, particularly compactness, will result in TVs and radios with antennas embedded inside the device itself.

### Automotive:

i. **Connected Vehicles**: taking advantage of the novel antennas advantages over ordinary antennas in use today enhances vehicle-to-everything (V2X) communications, crucial for the safety features in autonomous and semi-autonomous vehicles.

ii. **GPS Systems and Radar-Based Driver Assistance**: adopting the novel antenna in this technologies improves the accuracy and reliability of GPS navigation. Radar systems for driver assistance, like adaptive cruise control and collision avoidance can benefit from this technology resulting in more accurate data from the radar sensors.

### Healthcare:

i. **Wireless Monitoring Devices**: Vital for patient monitoring systems that require reliable wireless communication to transmit health data in real-time. With the adoption of the technology for health monitoring even outside of health facilities, increasing the range and enhancing the data transfer rates of these technologies greatly enhances their operation. For compact devices that maybe be implanted in the patient's body, the compact nature of our antenna and the higher efficiency will greatly enhance the versatility of the devices in this sector

ii. **Medical Imaging Systems**: with the advent of portable imaging devices our novel antenna can potentially enhance the wireless capabilities of these devices, making medical diagnostics more accessible. Imaging technologies that make use of radio waves like MRI can also benefit from this novel technology by the invention of more versatile transmitters and also more sensitive receivers leading to enhanced imaging.

### Aerospace and Defense:

i. **Radar Systems**: The technology's wide bandwidth and high efficiency can improve radar system performance. Radar systems are crucial for surveillance and navigation. This technology will also encourage the design of much smaller radar antennas particularly lower frequency radar systems.

ii. **Navigation Systems**: Adoption of this antenna technology in this sector will lead to design of systems that Support advanced navigation and communication tools in aircraft and spacecraft, offering precise data transmission and reception. A good example is underwater communication systems used by submarines and other under water vehicles, due to higher levels of attenuation by water (as compared to air), only very low frequency signals, often in khz can penetrate water efficiently enough for undersea communication to be effective. The antennas used for this kind of communication tend to be really bulky often being placed in large fields due to the size of individual antenna elements. Our proposed antenna eliminates the dependence of antenna efficiency of operation on the size of the antenna as compared to the target signals wavelength. This will result in compact antennas for these systems comparable in size to those in use in higher frequency applications.

### Public Safety:

i. **Emergency Communication Systems**: Ensures reliable communication for first responders, supporting systems like public safety LTE networks. By adopting our technology more efficient and reliable communication systems can be achieved in this sector. This would be greatly appreciated in this field considering that people in this sector work under conditions that might be unfavorable for other civilian communication systems.

ii. **Disaster Recovery**: Plays a key role in deployable communication systems used for disaster recovery efforts, where rapid setup and reliable performance are paramount. Most of the time conditions encountered in this sector call for communication systems that are reliable, versatile and adaptable to the difficult conditions. The advantages afforded by our antenna will be of great use in this sector.

### Research and Scientific Applications:

i. **Radio Astronomy**: The antennas in this sector require sensitivities that are unmatched; this results in very bulky and expensive systems that are not accessible even to most astronomers.

The advantages offered by our antenna in terms of size, efficiency, sensitivity and an ultrabroad bandwidth enhances the capabilities of radio telescopes, enabling cheaper more accessible systems resulting in more detailed observations of the universe.

ii. **Environmental Monitoring:** Wireless sensor networks for monitoring environmental parameters mostly involves remote systems making measurements of different parameters and sending the information gathered wirelessly, contributing to research on climate change and natural disaster prediction. These systems require to be very efficient since power might not be readily available in the remote locations they operate from.

The Single Layer Capacitor-Based Antenna Technology offers transformative potential across multiple sectors, providing the foundational support for next-generation wireless applications. By addressing current challenges and unlocking new possibilities, it paves the way for innovative solutions that will shape the future of connectivity.

## 6. Integration and Implementation

## Streamlining the Adoption of Single Layer Capacitor-Based Antenna Technology

Adopting the Single Layer Capacitor-Based Antenna Technology entails understanding its compatibility with existing systems and the steps required for seamless integration. From research the one compatibility issue that was identified is impedance mismatch. From the technical data analysis section, around the resonant frequency of a single layer capacitor, the impedance of the capacitor remains very low although it increases steadily as you move away from the resonant frequency. This gives the single layer capacitor based antenna its wide bandwidth. However, the antenna input/output pins of most modern communication systems are adapted to 50ohms impedance. To integrate our antenna into existing systems an impedance matching mechanism becomes necessary. Due to the low reactive resistance of the capacitor around the resonant frequency, a purely resistive impedance matching network can be implemented as seen in the technical data section (see Figures 8 and 9). This is the simplest impedance matching network for this type of antenna. For systems requiring narrower bandwidth, a reactive tuning circuit can be added to the system to control the bandwidth. Though, there is no design constrain that prevents engineers from designing lower impedance antenna inputs other than the fact that traditional antennas have impedances around this value. With a new antenna presenting a different set of proprieties, the design space opens up further. As engineers design their systems to take full advantage of the unique properties of this antenna more versatile designs around antenna input properties will be adopted. This section offers a general guidance for manufacturers, engineers, and designers looking to harness the technology's benefits across different applications.

**Integration Steps:**

i. **Assessment of Requirements:** Begin by evaluating the specific needs of your device or system, including frequency range, power handling, and size constraints, to determine how the technology can best meet these requirements.

ii. **Selection of Compatible Capacitors:** Based on the assessment, select the appropriate single-layer capacitors that meet the technical specifications, such as frequency response and power handling capabilities. A wide variety of single layer capacitors from a large number of manufacturers are already available in the market. In our research we have developed simulation tools that can be used for this task.

iii. **Impedance Matching:** Implement an impedance matching circuit to ensure maximum power transfer between the antenna and the transceiver, enhancing the overall system performance. A tuning circuit can also be implemented to control the bandwidth according to your requirements.

iv. **Design Adaptation:** Modify the existing PCB design to accommodate the single-layer capacitors and any support circuit that maybe included, taking into consideration the layout and spacing to optimize performance and efficiency.

v. **Prototype Testing:** Develop a prototype incorporating the antenna technology. Conduct comprehensive testing to evaluate performance, identify potential issues, and make necessary adjustments.

**Key Considerations for Implementation:**

i. **Compatibility with Existing Systems:** The technology's versatile frequency range and compact design make it compatible with a wide range of devices and communication standards.

ii. **Efficiency Optimization:** Leverage the high efficiency of the antenna to improve battery life in portable devices or to enhance signal strength in fixed installations.

iii. **Cost-Effectiveness:** Take advantage of the technology's cost-saving potential by evaluating the total cost of implementation, including materials, manufacturing, and potential savings from reduced size and enhanced performance.

**Support and Resources:** Manufacturers and developers looking to adopt this technology can access a range of support services and resources:

i. **Technical Documentation:** Detailed description of this technology including the simulations is available, design guidelines, and integration tips are also available to ensure a smooth adaptation process.
ii. **Expert Consultation:** Our team of technical experts is available to provide advice, answer questions, and assist with design challenges.
iii. **Community Forums:** Join discussions with other professionals who have successfully integrated the technology into their projects, sharing insights and solutions.

Integrating Single Layer Capacitor-Based Antenna Technology into your devices or systems represents a forward-thinking approach to meeting the demands of modern wireless communication. By following these guidelines and leveraging available resources, you can unlock the full potential of this innovative technology, driving enhanced connectivity and performance in your applications.

## 7. Conclusion and Future Outlook

### Envisioning a Connected World Enhanced by Single Layer Capacitor-Based Antenna Technology

As we stand on the brink of a new era in wireless communication, the Single Layer Capacitor-Based Antenna Technology emerges not just as an innovation but as a pivotal force capable of redefining the landscape of connectivity. Through its compact size, high efficiency, broad frequency range, and cost-effectiveness, this technology offers a promising solution to the pressing challenges of modern wireless systems. It stands as a testament to the power of ingenuity in meeting the demands of an increasingly digital and interconnected world.

**Achievements and Impact:** The journey of this technology from conceptualization to realization underscores a significant leap in antenna design, showcasing the potential to enhance a wide array of applications across telecommunications, consumer electronics, automotive, healthcare, aerospace and defense, public safety, and research and scientific applications. By providing a versatile, efficient, and scalable solution, it enables advancements that were once constrained by the limitations of traditional antenna technologies.

**Looking Ahead:** The future of Single Layer Capacitor-Based Antenna Technology is not just about the continuation of its current trajectory but about the exploration of new frontiers. The adaptability and performance of this technology open up possibilities for its integration into emerging fields such as 6G networks, quantum computing communications, and beyond. Research and development efforts will continue to refine and expand its capabilities, potentially unlocking even greater efficiencies and supporting the evolution of even more compact and powerful wireless devices.

Furthermore, as the global demand for seamless and robust connectivity grows, the role of Single Layer Capacitor-Based Antenna Technology in enabling smart cities, advanced IoT ecosystems, and next-generation mobile networks will become increasingly central. Collaborations between academia, industry, and regulatory bodies will be crucial in navigating the challenges of standardization, interoperability, and environmental sustainability.

**Embracing the Future:** As we look forward, the promise of Single Layer Capacitor-Based Antenna Technology in shaping the future of connectivity is clear. Its continued development and adoption will not only drive technological innovation but also foster a more connected, efficient, and sustainable world. We stand at the cusp of unlocking new capabilities that will propel us towards an even more interconnected future, with Single Layer Capacitor-Based Antenna Technology leading the way.

**Call to action:** While our current presentation of the Single Layer Capacitor-Based Antenna Technology is grounded in solid theoretical foundations and preliminary testing results, we recognize the limitations imposed by our current equipment capabilities. As such, detailed comparative performance metrics and deeper technical insights are areas we aim to expand upon through collaborative research and development. As we look towards the future, we are eager to collaborate with industry leaders, research institutions, and technology innovators to conduct the in-depth studies and advanced experimentation that our technology deserves. Our vision extends beyond the current capabilities, aiming to leverage partnerships to access better equipment, expertise, and funding. These collaborations are crucial for overcoming the existing limitations and for ensuring that our technology can be rigorously tested and optimized for a wide range of applications. We are committed to transparency and mutual growth with our partners, and we believe that together, we can achieve remarkable strides in making Single Layer Capacitor-Based Antenna Technology a cornerstone of next-generation wireless communication systems.

## Acknowledgement

We would wish to most sincerely thank the Technical university of Kenya for offering the motivation on technological innovations in which this work is grounded and workplace environment. I particularly acknowledge the support-technical and otherwise- offered by the Director School of Physics and Earth Sciences Professor George Amolo, Lecturer Dr. Lydiah Gachahi of the Department of Geosciences and the Environment, I would also like to take this opportunity to acknowledge the support and contribution of Mr. Robert Mwangi a colleague technician in the physics laboratories of the TUK.

**Bibliography and Further Reading**

To ensure the highest degree of accuracy and reliability in the discussions presented in this white paper, all technological insights, market analyses, and future outlooks are grounded in comprehensive research, including patents, scientific literature, and industry reports. While specific references used in the drafting of this document are generalized and not directly cited, the following sources are recommended for those interested in exploring the concepts, technologies, and market potentials discussed:

**Industry Standards and Protocols for Wireless Communication:**

- **3GPP Technical Specifications:** https://www.3gpp.org/specifications-technologies

- **Institute of Electrical and Electronics Engineers 802Standards:** https://www.ieee802.org/

- **European Telecommunications Standards Institute (ETSI) Standards:** https://www.etsi.org/

**Patents Relating to Antenna Design and Wireless Communication Technologies:**

- **United States Patent and Trademark Office (USPTO):** https://www.uspto.gov/

- **European Patent Office (EPO):** https://worldwide.espacenet.com/

**Scientific Journals and Conference Proceedings on Wireless Technologies and Electromagnetic Theory:**

- **IEEE transactions on antennas and propagation:**

  https://ieeexplore.ieee.org/xpl/RecentIssue.jsp?punumber=8

- **IEEE Antennas and Wireless Propagation Letters:**

  https://ieeexplore.ieee.org/xpl/RecentIssue.jsp?punumber=7727

- **Progress In Electromagnetics Research:** https://www.jpier.org/

**Market Research Reports on Telecommunications and Consumer Electronics:**

- **Gartner:** https://www.gartner.com/en

- **IDC:** https://www.idc.com/

- **Statista:** https://www.statista.com/

**Books and Textbooks on Antenna Design and RF Engineering:**

- **Antenna Theory and Design by John D. Kraus and Ronald J. Marhefka**

- **Antenna Engineering Handbook by Robert C. Johnson, Harold B. Crawford and W. Jasik**

- **RF and Microwave Engineering by Matthew M. Radmanesh**

**Technical Blogs and Forums on Wireless Communication:**

- **IEEE Spectrum: https://www.ieee.org/**

- **EDN Network: https://ednnetwork.org/**

- **RF Cafe: https://www.rfcafe.com/**

**Regulatory and Standards Bodies Documentation:**

- **Federal Communications Commission (FCC): https://www.fcc.gov/**

- **International Telecommunication Union (ITU): https://www.itu.int/**

**Further Reading:** For those seeking to deepen their understanding of Single Layer Capacitor-Based Antenna Technology and its applications, the following topics are recommended for further exploration:

i. Advances in single layer Capacitor Materials and Fabrication Techniques

ii. The Evolution of Wireless Communication Standards from 5G to 6G and Beyond

iii. Environmental and Health Considerations in Antenna Design and Deployment

iv. Case Studies on the Integration of Advanced Antenna Technologies in Real-World Applications

**List of Figures:**

**Figure 1:** The radiation pattern of a single element of the novel antenna showing the four main lobes.

**Figure 2:** The nearfield radiation pattern of an array made up of the Novel Antenna elements, showing the four main lobes and also the minor lobes.

**Figure 3:** The far field radiation pattern of an array made up of Novel Antenna elements showing the four main lobes.

**Figure 4:** The s-parameter data visualization of-from top left moving clockwise- s11, s12, s21, and s22 parameters.

**Figure 5:** A graph of the frequency response of impedance, reactance, and resistance of the antenna.

**Figure 6:** A graph of the frequency response of the dissipative factor of the base element.

**Figure 7:** A graph of signal strength against time for the novel antenna(dataset2) and the micro strip antenna(dataset1)

**Figure 8:** A graph of impedance vs frequency for a 50ohm source matched to the novel antenna using a resistor.

**Figure 9:** A graph of VSWR vs frequency for a 50ohm source matched to the novel antenna using a resistor.